\documentclass[11pt]{emulateapj}
\usepackage{graphicx}
\usepackage{amssymb,amsmath}
\usepackage{natbib}

\begin{document}

\title{The Intrinsic Quasar Luminosity Function: Accounting for Accretion Disk Anisotropy}
\author{M.A. DiPompeo\altaffilmark{1}, A.D. Myers\altaffilmark{1}, M.S. Brotherton\altaffilmark{1}, J.C. Runnoe\altaffilmark{3}, R.F. Green\altaffilmark{2},}
\altaffiltext{1}{University of Wyoming, Dept. of Physics \& Astronomy 3905, 1000 E. University, Laramie, WY 82071, USA}
\altaffiltext{2}{Large Binocular Telescope Observatory, University of Arizona, 933 N. Cherry Avenue, Tucson, AZ 85721, USA}
\altaffiltext{3}{Penn State University, Dept. of Astronomy \& Astrophysics, 413 Davey Lab, University Park, PA 16802, USA}

\begin{abstract}
Quasar luminosity functions are a fundamental probe of the growth and evolution of supermassive black holes.  Measuring the intrinsic luminosity function is difficult in practice, due to a multitude of observational and systematic effects.  As sample sizes increase and measurement errors drop, characterizing the systematic effects is becoming more important.  It is well known that the continuum emission from the accretion disk of quasars is anisotropic --- in part due to its disk-like structure --- but current luminosity function calculations effectively assume isotropy over the range of unobscured lines of sight.  Here, we provide the first steps in characterizing the effect of random quasar orientations and simple models of anisotropy on observed luminosity functions.  We find that the effect of orientation is not insignificant and exceeds other potential corrections such as those from gravitational lensing of foreground structures. We argue that current observational constraints may overestimate the intrinsic luminosity function by as much as a factor of $\sim$2 on the bright end. This has implications for models of quasars and their role in the Universe, such as quasars' contribution to cosmological backgrounds. 

\end{abstract}


\section{INTRODUCTION}
The most luminous active galactic nuclei (AGN), generally referred to as quasars, are fundamental to the study of cosmology, the growth of structure, supermassive black hole - host galaxy co-evolution, and a multitude of other astrophysical processes.  Soon after the discovery of quasars, characterizing the quasar luminosity function (QLF) and its evolution with redshift became an area of intense study (e.g.\ Sandage 1961; Greenstein \& Matthews 1963; Burbidge 1967).  

Current theories indicate that the bulk of supermassive black holes (SMBHs) gain the majority of their mass via radiatively efficient accretion at high, quasar-level luminosity ($L_{bol} > 10^{45} \textrm{ erg s}^{-1}$; e.g.\ Salpeter 1964; Soltan 1982).  A measurement of the quasar LF thus provides constraints on the growth history of SMBHs (e.g.\ Madau \& Rees 2001; Volonteri et al.\ 2003; Volonteri \& Rees 2006).  Quasars are also likely key contributors to various cosmic backgrounds (e.g.\ the X-ray, UV and IR backgrounds; Henry 1991; Shanks et al.\ 1991; Madau 1992; Ueda et al.\ 2003; Dole et al.\ 2006; Hickox \& Markevitch 2006).  Finally, AGN and quasars likely play a key role in the evolution of galaxies, as the SMBH and host evolve in lock-step (e.g.\ the $M$-$\sigma$ relation; Ferrarese \& Merritt 2000; Gebhardt 2000).  Understanding the evolution of quasars sheds light on the complex nature of this co-evolution.

Accurately constraining the QLF is difficult, but measurements have been steadily improving as large surveys uncover more quasars (e.g.\ Oke 1983; Boyle et al.\ 1991; Hewett et al.\ 1993; Koehler et al.\ 1997; Goldschmidt \& Miller 1998; Boyle et al.\ 2000; Fan et al.\ 2001; Wolf et al.\ 2003; Croom et al.\ 2004; Richards et al.\ 2005; Jiang et al.\ 2006; Fontanot et al.\ 2007; Jiang et al.\ 2008; Croom et al.\ 2009; Willott et al.\ 2010; Glikman et al.\ 2011; Masters et al. 2012; Ross et al.\ 2013).  Many of the difficulties in measuring the QLF are related to the surveys required to find quasars in large numbers --- corrections must be applied for morphological incompleteness (real quasars discarded on morphological grounds), targeting incompleteness (quasars missed by target selection algorithms), and survey coverage completeness (e.g.\ Ross et al.\ 2013).  Several corrections for these factors must be applied to estimate the ``true'' QLF, introducing scatter between individual studies.

It is important to recognize that even this ``true'' QLF is an observed QLF, and not the intrinsic QLF, because of additional factors not usually corrected for in typical quasars.  One of these is orientation (e.g.\ Antonucci 1993), which we will focus upon in this paper.  Corrections for orientation are already applied to the LFs of some special classes of objects, such as blazars (e.g.\ Ajello et al.\ 2012).  These corrections are largely due to the effects of relativistic beaming due to the jet-on orientations of these objects, as opposed to pure geometric effects, but the idea is the same.

Essentially all investigations determining QLFs assume isotropic emission in the optical/ultraviolet.   However, quasar structures, from the large scale radio jets (e.g.\ Barthel 1989), to obscuring dusty tori (e.g.\ Ma \& Wang 2013), to the accretion disks that emit the optical/ultraviolet continuum that dominates the bolometric luminosity, are apparently axisymmetric rather than spherical.    There is observational evidence to suggest that the continuum of type 1 quasars varies by a little over a factor of two due to orientation (e.g., Runnoe et al.\ 2013), which is roughly consistent with theoretical expectations (e.g.\ Laor \& Netzer 1989; Nemmen \& Brotherton 2010).  There is also some evidence that the X-rays are not isotropic either, and vary in a similar way to the UV continuum (Zhang 2005; Runnoe et al.\ 2013). Theoretical considerations include projection effects (a simple cos $\theta$\ term, where the angle $\theta$\ is the angle of the spin axis relative to the line of sight), as well as relativistic effects and limb darkening.  Moreover, the opening angle of any obscuring structures and their variation with luminosity (e.g.\ Lawrence 1991) or redshift will also affect the range of orientation effects.  Because QLFs steepen at higher luminosities, orientation effects will not be symmetric, and high luminosities will be preferentially occupied by more face-on quasars compared to lower luminosity objects (e.g.\ Urry et al.\ 1991)

While observed QLFs are fine for applications where the observed properties are the primary interest, such as predicting number counts in surveys, there are other applications for which they are inadequate. For instance, almost all simulations investigating mutual black hole, galaxy evolution do not take into consideration orientation issues and treat the observed QLFs as intrinsic (e.g.\ Hopkins et al.\ 2008, 2010), while we show here that it is not.

Here we launch an initial, quantitative investigation examining the difference between observed and intrinsic QLFs on the basis of orientation.  As a first step, we will only 
investigate simple and robust orientation effects to quantify the magnitude of the effect. 
In this paper we present models of the effects of orientation on the QLF, and present an intrinsic QLF with the effects of anisotropy removed.  We use a cosmology where $H_0 = 71$ km s$^{-1}$ Mpc$^{-1}$, $\Omega_M=0.27$, and $\Omega_{\Lambda} =0.73$ for all calculated parameters (Komatsu et al.\ 2011).

\section{METHODOLOGY}

\subsection{Anisotropy Models}

Unfortunately, there is no single, recognized, reliable description of the dependence of the quasar optical/ultraviolet continuum on viewing angle.  There is typically a recognition that geometrical projection, 
relativistic effects, and limb darkening are likely important, and that we understand the relevant physics of these at some level of reliability (e.g., Shakura \& Sunyaev 1973; Netzer 1985; Laor \& Netzer 1989; 
Hubeny et al. 2000; Fukue \& Akizuki 2006), 
although perhaps not always or robustly (e.g., Shang et al.\ 2005).  The reality is, however, that accretion disks in theory or reality 
do depend in detail on a number of parameters and their various influences depend on a number of factors. 

We look to the empirical findings of Runnoe et al. (2013), which show that type 1 quasars vary in their optical/ultraviolet continuum luminosity by factors of 2-3 over their range in viewing angles, for guidance.
Type 1 quasars are seen generally at angles less than $\sim$45 degrees, more ``face-on'' compared to random viewing angles.  Given an anisotropy based on projection effects alone, this indicates
that typical type 1 quasars are brighter than average, which will already bias quasar luminosities to be above average.  Additionally, limb darkening and relativistic affects are thought to be more 
important at larger viewing angles.  The accretion disk models of Hubeny et al. (2000), as examined by Nemmen \& Brotherton (2010), show that the higher luminosity quasars
should have an anisotropic luminosity pattern consistent with geometric projection within likely opening angles.  This is consistent with the factor of 2-3 shown empirically in Runnoe et al.\ (2013).

Therefore, for this initial investigation we assume only the simplest Newtonian disk model with a $\cos \theta$  dependence for the optical/ultraviolet continuum.  This term dominates over any other effects 
(e.g.\ limb darkening, relativity), except at the largest inclinations ($>85^{\circ}$; see Figure 3 of Nemmen \& Brotherton 2010).
 
\subsection{Observed LFs}
Our aim is to both characterize the effect of orientation on the measured  QLF, as well as compare the magnitude of the effect with our ability to accurately measure the LF due to other effects.  Therefore, we select two different LFs from the literature that have a range in observed parameters.  Taken together, these represent how accurately we are able to measure the  QLF overall.  These are simply chosen to represent a range of observed LFs --- the effects of orientation, as detailed and discussed further, will not change dramatically if we selected different luminosity functions.  Also, many newer studies focus on quasars at higher redshift (above $z\sim2.5$, e.g.\ Ross et al. 2013), while the quasar number density peaks below this.  The redshift range covered by the LFs used here is large enough for this first analysis, and despite the fact that the shape of the QLF likely evolves with redshift we argue below that this evolution has minor effects on the role of orientation.

\begin{figure}
\hspace{-0.1cm}
   \includegraphics[width=8cm]{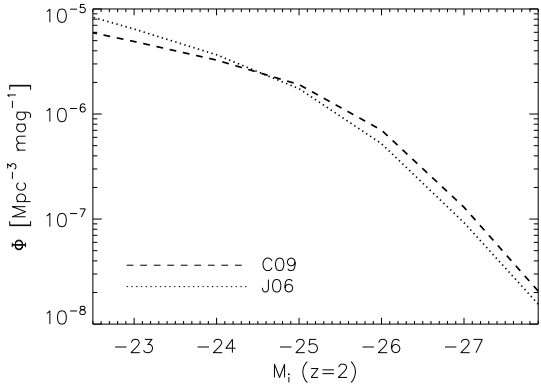}
  \caption{The two  QLFs (at $z=1.5$) used as a starting point for our simulations.  They have been adjusted to the $i$-band k-corrected to $z=2$, using the k-corrections and empirical relations of Richards et al.\ (2006) and C09.\label{fig:lit}}
\end{figure}

Both of the chosen QLFs are characterized by broken power-laws, though they use slightly different functional forms.  The selected LFs are shown in Figure~\ref{fig:lit} and are summarized below.  Both have been adjusted to $i$-band magnitudes k-corrected to $z=2$.  In all cases we will use $\alpha$ and $\beta$ as the bright- and faint-end LF slopes, respectively.

\vspace{0.3cm}

\noindent \textit{(1) Jiang et al.\ (2006; hereafter J06)}:  J06 fit their LF including a density evolution term with redshift:
\begin{equation}
\Phi (M,z) = \frac{\Phi(M^{*}) \rho_{D}(z)}{10^{0.4(\alpha + 1)(M-M^{*})}+10^{0.4(\beta+1)(M-M^{*})}},
\end{equation}
where $\rho_{D}(z) = 10^{-B(z-2)}$.  An exponential is used to describe the evolution of $M^{*}$:
\begin{equation}
M^{*}(z) = M^{*}(0) -1.08 k \tau
\end{equation}
where $\tau$ is the lookback time.
The LF fit parameters from J06 are given in the first column of Table~\ref{table:lit}.  Their results are presented for absolute magnitudes in the $g$-band k-corrected to $z=0$.  To convert this LF into $i$-band magnitudes at $z=2$, we apply k-corrections from Richards et al. (2006), including a correction for the emission-line flux, and assume a power-law continuum slope of $\alpha_{UV} = -0.5$.

\vspace{0.3cm}

\noindent \textit{(2) Croom et al.\ (2009; hereafter C09)}:  This LF is characterized as:
\begin{equation}
\Phi (M,z) = \frac{\Phi(M^{*})}{10^{0.4(\alpha + 1)(M-M^{*})}+10^{0.4(\beta+1)(M-M^{*})}}
\end{equation}
where $M^{*}$ is a function of $z$ such that
\begin{equation}
M^{*}(z) = M^{*}(0) - 2.5(k_1 z +k_2 z^2).
\end{equation}

The fit parameters are taken from the third row of Table 2 in C09, and listed in the second column of Table~\ref{table:lit}.  An absolute magnitude cut to be considered a quasar is imposed on their sample, at $M < -21.5$.  Their results are presented as $g$-band magnitudes k-corrected to $z=2$.  The k-correction we apply to this LF to make it directly comparable to J06 uses the relationships in Richards et al.\ (2006), but follows C09 by using a continuum slope of $\alpha_{UV} = -0.3$ and also uses their empirical determination of the contribution from emission lines.

\begin{deluxetable}{lccc}
 \tabletypesize{\scriptsize}
 \tablewidth{0pt}
 \tablecaption{Literature LF fit parameters\label{table:lit}}
 \tablehead{
                               &  \colhead{J06}                      &&  \colhead{C09}                            \\
                               & \colhead{$M_g$ ($z=0$)}  && \colhead{$M_g$ ($z=2$)}
    }
   \startdata
 $k$               &       7.5                                & &    \nodata                                  \\
$k_1$           &   \nodata                            & &        1.46                                    \\
$k_2$           &   \nodata                            & &        $-$0.33                              \\
$\alpha$      &    $-$3.25                           &&          $-$3.33                             \\
$\beta$        &    $-$1.55                           & &         $-$1.41                              \\
$M^{*}$        &    $-$19.5                           & &        $-$22.17                             \\
$\Phi^{*}$   &   $1.02 \times 10^{-6}$    & &        $1.67 \times 10^{-6}$       \\
$B$               &  0.45                                   & &           \nodata                              
  
   \enddata
   \tablecomments{The fit parameters for the J06, and C09 LFs, from their respective sources (see \S2.1).}
\end{deluxetable}

\subsection{Simulations}
Due to the anisotropic nature of quasar accretion disk emission, a given quasar would have a range of measured fluxes, and inferred luminosities, if it were seen from different viewing angles.  Even if every quasar had the same intrinsic luminosity, their random orientations would lead us to measure a distribution of luminosities if we naively assumed their emission was isotropic.  The only fair way to compare quasars or characterize their properties as a population is by correcting their fluxes to the value that would be measured if they were all seen from the same line of sight.  Failing to apply this viewing angle correction is implicitly assuming that the emission is isotropic, which we know is invalid.  Not only is it necessary to adjust the fluxes to what they would be if all quasars had the \textit{same} viewing angle, but \textit{which} angle we correct to is important.  In order to appropriately apply a conversion from apparent magnitudes (fluxes) to absolute magnitudes (luminosities), we need to adjust the observed fluxes to their value at the angle for which an observation of an isotropic and anisotropic emitter would be the same.  Here we detail our methods for applying this correction.

\begin{figure*}
 \hspace{2cm}
   \includegraphics[width=13cm]{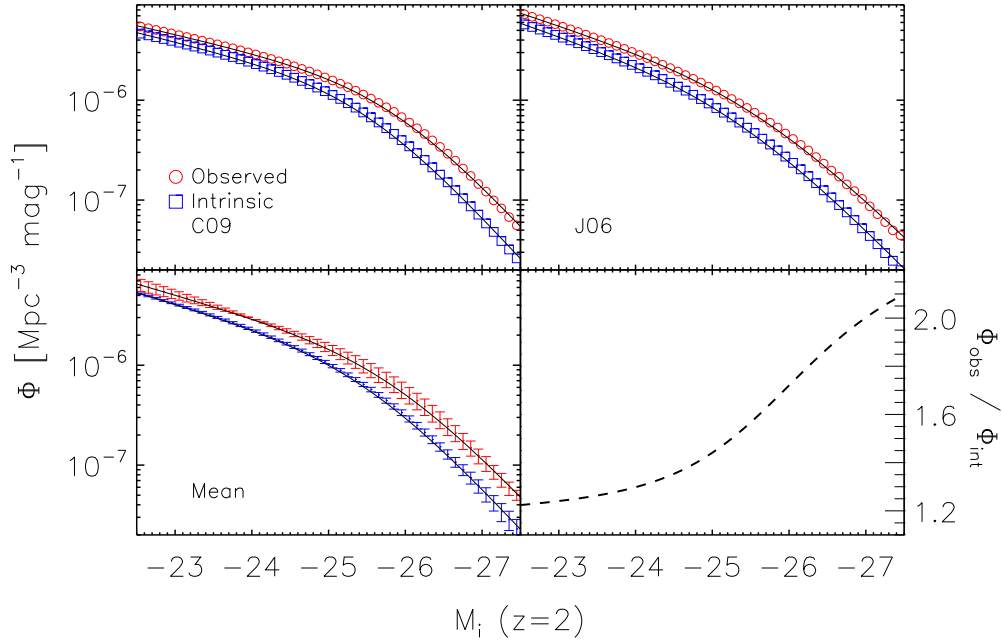}
  \caption{\textit{Top left, top right:} The observed and intrinsic (orientation corrected) individual quasar LFs. \textit{Bottom left:} The average of the top panels.  Error bars on the observed mean LF indicate scatter between different studies, while the error bars on the intrinsic LF indicate intrinsic scatter due to the fact that individual quasars have unknown, random viewing angles.  Black lines in all panels show our fits, with the parameters listed in Table~\ref{table:fits}.  \textit{Bottom right:} The residuals from the mean observed and intrinsic LFs.  The effect of orientation is significant, especially on the bright end.\label{fig:phi}}
\end{figure*}

Our first step is to generate a mock population of quasars using the observed luminosity functions.  To do this, we integrate each observed LF to generate distributions of $N$ quasars (per deg$^2$) as a function of $z$ and $M$ (from $0 < z < 5$; bins have widths of 0.1 for both parameters), including the evolution of the LFs with $z$.  The maximum magnitude correction due to orientation in our models (see below) will be $< 1$ mag, and we integrate over a range of $M$ sufficient to analyze the results at typical quasar luminosities, as well as allow objects to shift in or out of samples.  While our simulated range of $z$ is quite large to allow for analysis at any redshift, we present results averaged over the range $0.5 < z < 2.5$, a range typical of many real LF studies.  We find that the effect of orientation does not change significantly as the LF evolves with redshift, though our models can take this into account.  These generated luminosity functions averaged over $z$ are shown as the ``observed'' values (red circles) in the top panels of Figure~\ref{fig:phi}.

We then assign each quasar a random orientation consistent with an axis-symmetric geometry, where face-on orientations are less likely than edge-on (see the discussion section for possible factors that can influence this distribution).  We limit the available viewing angles to $\theta < 60^{\circ}$, assuming that this is the maximum angle before the ``dusty torus'' obscures the nuclear regions.  This will provide a conservative limiting case, as discussed in \S3.  We will also address the role of changing opening angles, as well as receding torus models in \S3.  

Each quasar is assigned a random luminosity ($M$; uniformly within its bin).  To get to the intrinsic value of $M$, and thus the intrinsic QLF, we must adjust the actual magnitude to what it would be \textit{at an angle where the conversion from apparent to absolute magnitude is the same in the isotropic and anisotropic cases}.  Again, in these simulations we only consider the simplest Newtonian disk model, which is a $\cos \theta$ correction.  To find the angle to correct to, we equate the total flux from an accretion disk considering geometric effects to the flux measured assuming isotropy:
\begin{equation}
2F_0 \int_{0}^{\frac{\pi}{2}} \cos \theta d\theta \int_{0}^{2\pi} d\phi = 4 \pi F_0 \cos \theta
\end{equation}
where $F_0$ is the face-on flux, and $\theta$ is the viewing angle.  The value of $\theta$ for which these expressions are equivalent is the critical angle ($\theta_c$) at which the observed flux assuming isotropy approximates the intrinsic total flux.  Solving gives $\theta_c = 60^{\circ}$.  Therefore, we take each assigned (``observed'') $M$ and $\theta$, and correct it to the magnitude that would be observed if the viewing angle were $\theta = 60^{\circ}$.  Because 60$^{\circ}$ is the maximum allowed viewing angle, all absolute magnitudes are adjusted to their value at a larger angle and therefore \textit{fainter} than the observed values.  

We then re-bin the number counts in $z$ and $M$, repeat the process 1000 times, and average the resulting luminosity functions from each iteration to find the average effect of orientation.  This allows a comparison between the observed and intrinsic luminosity function.  Again, intrinsic in this case means the LF if all quasars were viewed at the same angle, $\theta = 60^{\circ}$.  These results are shown as blue squares in the top panels of Figure~\ref{fig:phi}.

The difference between the C09 and J06 observed LFs represents the limits on our ability to measure it accurately (assuming both methods to calculate it are equally valid).  The third (bottom left) panel of Figure~\ref{fig:phi} shows an average of the two observed LFs. We can compare the magnitude of the orientation correction to the magnitude of the difference between the observed values, which is shown in the error bars in this panel.  We then estimate the scatter on the intrinsic LF due to random orientations by taking the standard deviation of the number in each bin from the 1000 iterations of the simulations.  The orientation correction is much larger than the scatter between current observed LFs and the intrinsic scatter due to random orientations.  The final panel shows the residuals of the average observed and intrinsic LFs to illustrate the magnitude of the effect.

The method outlined above does not \textit{necessarily} create the intrinsic LF, as the output of the simulation depends on the slope of the input LF.  In other words, if the correction from orientation sufficiently alters the slope of the LF, running the simulation one way (adjusting all objects to the same viewing angle) and then back (re-randomizing the viewing angles) will not produce the original LF.  Additionally, the observed luminosity is a function of viewing angle (brighter objects are more likely to be face-on), which alters the probability distribution of viewing angles of observed objects that our simulation starts with.  Therefore, we re-randomize the viewing angles and calculate the observed LFs from our simulation, and this does indeed reproduce the actual observed LFs we start from within a few percent on the bright end and within a percent at the faint end.  This small difference is well within the scatter due to random orientations, and even within the scatter between observed LFs.  Therefore, our results are fully consistent with the intrinsic LFs, with no need to further tweak the correction via a fully forward-modeled simulation.

In order to quantify these results and accurately compare them, we perform our own fits to the observed and intrinsic LFs.  For consistency, in all cases we use the power-law parameterization given in equations 3 and 4.  We hold $k_1$ and $k_2$ constant in our fits; these are not particularly well constrained, and do not have a large effect on the fits.  The other parameters are more important for analyzing the effect of orientation on the observed LF.  Therefore, we simply use the values from C09, $k_1=1.46$ and $k_2=-0.33$.   Note that because we are averaging the LF over a large range of redshifts before performing our fits, the fits to the observed LFs appear different from those in the original references.  However, if we perform our fits on the LF at a given redshift, they agree with those used as inputs (at least in the case of C09, since we use their parametrization).  The fits are shown as black lines in the panels of Figure~\ref{fig:phi}, and the parameters are summarized in Table~\ref{table:fits}.  Note that the final row, labeled as ``mean'', is a fit to the averaged QLFs, not simply an average of the parameters from the individual fits. We also perform the fits on each individual iteration that randomizes the orientations, and take the standard deviation of each fit parameter as the systematic scatter in the LF parameters due to the fact that we don't know the orientation of individual objects --- this scatter is given in the final four columns of Table~\ref{table:fits}.

\begin{deluxetable*}{lccccccccccccccc}
 \tabletypesize{\scriptsize}
 \tablewidth{0pt}
 \tablecaption{Fit parameters for the observed/intrinsic luminosity function\label{table:fits}}
 \tablehead{
                                 &            \multicolumn{4}{c}{Observed}    &        &    \multicolumn{4}{c}{Intrinsic}   &  &   \multicolumn{4}{c}{Scatter due to orientation} \\
                                                          \cline{2-5}                                                      \cline{7-10}                                            \cline{12-15}                          \\
    LF                        & \colhead {$\alpha$}  &  \colhead{$\beta$}  &  \colhead{$M^{*}$}  &  \colhead{$\Phi^{*} (\times 10^{-6})$}  & &  \colhead {$\alpha$}  &  \colhead{$\beta$}  &  \colhead{$M^{*}(0)$}  &   \colhead{$\Phi^{*} (\times 10^{-6})$}   & &  \colhead {$\alpha$}  &  \colhead{$\beta$}  &  \colhead{$M^{*}$}  &  \colhead{$\Phi^{*} (\times 10^{-6})$}
    }
   \startdata
   J06	&  $-$2.93  &  $-$1.61  & $-$22.09  & 1.30  & & $-$3.00  & $-$1.64  & $-$21.82  & 1.11  & & 0.12 & 0.07 & 0.18 & 0.26  \\ 
   C09	&  $-$3.11  &  $-$1.44  & $-$22.31  & 1.44  & & $-$3.13  & $-$1.46  &  $-$21.96 & 1.35  & & 0.07 & 0.02 & 0.07 & 0.10  \\
   Mean     &  $-$3.11  &  $-$1.48  & $-$22.48  & 1.07  & & $-$3.14  & $-$1.49  &  $-$22.13 & 1.01  & & 0.13 & 0.07 & 0.19 & 0.28   
   \enddata
   \tablecomments{These fits use the parametrization of equations 3 and 4, holding $k_1=1.46$ and $k_2=-0.33$.  The final row, the mean, is a fit to the average of the LFs, not an average of the individual fit parameters.  The final four columns give the scatter in the fit parameters from each iteration of the simulations, and represent how well these parameters can be constrained in samples where individual viewing angles are unknown.}
\end{deluxetable*}

\section{RESULTS \& DISCUSSION}
It is clear from ~\ref{fig:phi} that the intrinsic, orientation-corrected QLF falls well outside of the scatter in recent measurements.  The correction is largest on the bright end, where the slope of the QLF is steeper.  At the brightest luminosities, the intrinsic QLF is a factor of 2 or more smaller than what is observed.  Comparing the values in Table~\ref{table:fits} shows that the changes are systematic in nature --- the slopes at both the faint and bright ends decrease, and both $M^{*}$ and $\Phi^{*}$ increase, in all cases.  Aside from the overall magnitude of the average affect of orientation, the final four columns of Table~\ref{table:fits} indicate that we cannot constrain the shape of the quasar LF better than at best a few percent without knowing   the orientations of individual objects in a sample.

\subsection{Additional factors}
Our models have assumed that all quasars can be seen from a viewing angle range of $0-60^{\circ}$.  However, not all quasars are likely to have the same covering fraction, or opening angle of the obscuring torus.  It has been hypothesized that this is a function of the quasar luminosity (Lawrence 1991), and observed in several studies (e.g.\ Arshakian 2005; Cleary et al.\ 2007; Calderone et al.\ 2012).  Our choice of $60^{\circ}$ as the maximum allowed opening angle is likely more of an upper limit.  Allowing for a varying opening angle will cause an enhancement of our results --- if the maximum viewing angle allowed is less than 60 degrees, then more objects will have a larger correction (be made fainter) to what their observed magnitude would be at $60^{\circ}$.  The results of these simulations can thus be seen as the \textit{minimum} effect due to orientation.  This is enhanced by the fact that we have left out other sources of anisotropy and focused on the geometric correction only --- the addition of other effects will also enhance the change.

There are several other factors that can come into play when considering the probability of finding a quasar at a given viewing angle ($P(\theta)$), in addition to pure geometric effects, which could be used in the future to develop more complicated models.  For example, the most luminous quasars are probably more likely to be found in a face-on orientation.  In fact, $P(\theta)$ is probably also a function of many complicated and (in some cases) poorly constrained parameters for individual objects --- redshift, black hole mass, accretion disk wind strength/opening angle, cloud distribution, and dusty torus size to name a few.  However, geometric projection is one of the largest effects, and the work here provides a first-order correction.

\subsection{Comparison with other systematics}
How do the effects of orientation compare with other possible effects that change the intrinsic QLF, outside of observational constraints?  As an example, we compared the magnitude of orientation effects with another potential mechanism that can alter the observed QLF --- foreground lensing of quasars.  The observed number densities (and thus LFs) of distant objects can be altered as their light passes through foreground structures, like galaxy clusters (e.g.\ Marri et al.\ 2000).  Two effects come into play  --- a change in magnitude, and a change in effective area (and thus volume).  A summary of these effects (and their mathematical form) can be found in the appendix of Myers et al.\ (2003).  Our model below assumes a singular isothermal sphere (SIS) model for the dark matter in clusters.

To simulate the role of lensing, we start in the same manner as outlined above, with a distribution of quasar numbers as a function of redshift and magnitude.  We assign each quasar in a bin a position and redshift from a random quasar in the SDSS quasar catalog of Schneider et al.\ (2010).  Using the SDSS MaxBCG catalog of Koester et al.\ (2007), a catalog of 13,823 galaxy clusters from SDSS, we find the distance of each simulated quasar from the center of the nearest cluster.  A key parameter in the lensing equations is the velocity dispersion of the dark matter in the halo, which depends on the number of galaxies in the cluster.  We assign a dispersion consistent with the number of galaxies in the nearest cluster (see Figure 11 of Kloester et al.\ 2007).  We find the average magnification in each magnitude bin, de-magnify each observed magnitude, and also apply the net enhancement factor to account for the effect of the changing area in each bin (using a faint-end number density slope of $-0.1$).  The numbers are re-binned, and a lensing-corrected number density and QLF are calculated.  We find that this correction is quite a bit smaller than that due to orientation, with a maximum change at the bright end of about a factor of $\sim$1.1.

\subsection{Applications of this correction}
The final panel of Figure~\ref{fig:phi} illustrates a first-order correction to the QLF for orientation effects.  For observational studies, for example when predicting the number of quasars expected in a given survey, clearly the observed number densities and LFs are what matters.  But for models of quasar evolution and their role in the evolution of the Universe, we need an accurate representation of the intrinsic QLF.  For example, overestimating the number of the most luminous quasars will have implications for their contribution to various cosmological backgrounds and reionization.  Of course, the obscured population (e.g.\ ``type 2'' quasars) are already accounted for in these analyses via their observed number densities --- this work provides a starting point for moving analysis from the observed to the intrinsic QLF over all lines of sight to unobscured objects.

\section{CONCLUSIONS \& SUMMARY}
Using two different observed LFs and applying a simple $\cos \theta$ correction for anisotropic accretion disk emission, we have illustrated a first step in moving from observed to intrinsic quasar luminosity functions.  Neglecting anisotropy leads to an over-estimate of the QLF (and thus quasar number density), by as much as a factor of two at the bright end.  This estimate is conservative, as we have allowed all quasars to have a wide range of viewing angles, when some are likely to have larger covering fractions.  Additionally, we have applied the simplest Newtonian correction --- more complex anisotropy models will enhance the effects seen here.  Nevertheless, the corrections we derive exceed other effects such as magnification bias from gravitational lensing.  Using a typical broken power-law to fit our results, we see that orientation tends to decrease both the bright and faint-end slopes, while increasing the break luminosity.  We also show that the scatter introduced into the quasar LF from unknown individual orientations intrinsically limits our ability to measure the QLF to within a few percent.

When using the quasar luminosity function to predict the number of quasars a survey will recover, the observed values are sufficient.  However, in order to fully understand the cosmological evolution of quasars and their role in the evolution of the Universe, we must move from observed to intrinsic LFs.  As our ability to measure the QLF improves, these systematic effects will become more important to consider.  Additionally, as our knowledge of the true form of the emission anisotropies increases, as well as our understanding of their (possible) dependance on luminosity, these models can be refined to provide even better measures of the intrinsic quasar luminosity function.

\acknowledgements
MAD and ADM were partially supported by NASA through ADAP award NNX12AE38G and EPSCoR award NNX11AM18A and by the National Science Foundation through grant number 1211112.

\clearpage

\end{document}